\documentclass{article}
\usepackage{amsmath}
\usepackage{amsfonts}
\usepackage{amssymb}
\usepackage{graphicx}

\input{tcilatex}

\begin{document}

\title{Questions, Relevance and Relative Entropy\thanks{%
Presented at MaxEnt 2004, the 24th International Workshop on Bayesian
Inference and Maximum Entropy Methods (July 25-30, 2004, Garching bei
Munchen, Germany).}}
\author{Ariel Caticha \\
%EndAName
{\small Department of Physics, University at Albany-SUNY, }\\
{\small Albany, NY 12222, USA.\thanks{%
E-mail: ariel@albany.edu}}}
\date{}
\maketitle

\begin{abstract}
What is a question? According to Cox a question can be identified with the
set of assertions that constitute possible answers. In this paper we propose
a different approach that combines the notion that questions are requests
for information with the notion that probability distributions represent
uncertainties resulting from lack of information. This suggests that to each
probability distribution one can naturally associate that particular
question which requests the information that is missing and vice-versa. We
propose to represent questions $q$ by probability distributions Next we
consider how questions relate to each other: to what extent is finding the
answer to one question relevant to answering another? A natural measure of
relevance is derived by requiring that it satisfy three desirable features
(three axioms). We find that the relevance of a question $q$ to another
question $Q$ turns out to be the relative entropy $S[q,Q]$ of the
corresponding distributions. An application to statistical physics is
briefly considered.
\end{abstract}

\section{Introduction}

What is a question? According to Cox a question can be identified with the
set of assertions that constitute possible answers \cite{Cox61, Cox79}. A
number of applications and formal developments by R. Fry \cite{Fry95} and by
K. Knuth \cite{Knuth02} have proved that this definition is, indeed,
fruitful. It allows the elegant and powerful mathematics of the theory of
partially ordered sets to be brought to bear on the field of logical inquiry.

It is undeniable that Cox's simple definition captures an essential part of
the notion of question, but it is not clear that it completely exhausts it.
In this paper I approach the subject from a different direction. I turn to
an older definition according to which a question is a request for
information. From this point of view one should expect natural connections
between a theory of inquiry and a theory of information and to inductive
logic, that is, to reasoning in situations of uncertainty. In other words,
we should expect close connections to the notions of probability and of
entropy.

The problem with defining questions as requests for information is that the
definition is somewhat vague and it might not be clear how it could provide
a quantitative foundation for a theory of inquiry. We need to be more
specific about what we mean by a request for information. The necessary
refinement is suggested by the fact that a probability distribution is an
expression of doubt; it is an admission of uncertainty and, implicitly, a
request for information. We propose that a \emph{question is a probability
distribution}. In our approach a question goes beyond the simple enumeration
of the possible answers; it also reflects the specific information we
request. One of the purposes of this paper is to motivate this definition,
to show that it is useful and that the resulting theory does not contradict
but rather complements and extends the theory of inquiry based on the Cox
definition (section 2). Among other advantages it allows a straightforward
extension of the theory of logical inquiry to the continuum.

Incidentally, the realization that there exists a close relation between a
question and a probability distribution, a relation that for the purposes of
this paper we take to be identity, explains the remarkable and puzzling
\textquotedblleft neglect accorded to questions throughout most, if not all,
of the history of logic and probability\textquotedblright\ \cite{Cox79}.
There has been no urgent need to develop a theory that is explicitly a
theory of questions because many of the problems such a theory would solve
are already being solved by the theory of probability. \emph{The theory of
probability is implicitly a theory of inquiry}.

In section 3 we consider how questions relate to each other, more
specifically, we seek a quantitative measure of the extent to which
answering one question is relevant to answering another. This allows us to
rank questions according to how relevant they are to a given \emph{prior}
question. Three natural axioms are proposed that define a unique measure of
relevance. The axioms reflect the conviction that of all questions $q$
within a certain family, the most relevant to answering the prior question $%
Q $ is the $q$ that is \textquotedblleft closest\textquotedblright\ to $Q$
in that it requests \textquotedblleft nearly\textquotedblright\ the same
information. Perhaps it is not surprising that once we have identified
questions with probability distributions we find that the relevance of
question $q$ to question $Q$ turns out to be the relative entropy $S[q,Q]$
of the corresponding distributions.

A possible objection is that many of the mathematical equations in this
paper have previously appeared elsewhere \cite{Caticha04} and all we have
done is change the names assigned to the various symbols: what used to be
called a probability distribution is now called a question, what was
previously called an amount of information or a degree of preference is now
called relevance; in essence this paper is nothing more than mere word
games. To a certain extent this is, of course, true: we are playing word
games, but these are not \textquotedblleft mere\textquotedblright\ word
games. It is precisely the assigning of different meanings to the old
symbols that suggests how to use them in new ways and allows one to tackle
new problems. One familiar example is Bayes' theorem: both frequentists and
Bayesians can prove the theorem but it is only under a Bayesian
interpretation that its full potential can be recognized and exploited.

The present paper represents one more step in an ongoing program to extend
the range of applicability of the concept of entropy: here we interpret
entropy as a measure of relevance and use it as a tool for inquiry.
Shannon's use of entropy as a measure of amount of information was exploited
by Jaynes to develop a general method to assign probabilities -- the MaxEnt
method. Later work by Shore and Johnson, Skilling, and others culminated in
the realization that entropy is a tool for processing information, for
updating from a prior distribution to a posterior distribution when
information is supplied in the form of constraints \cite{Caticha04,
ShoreJohnson80, Skilling88}. When used in this way entropy needs no
interpretation, we do not need to know what it means, we just need to know
how to use it. This form of the maximum entropy method, the ME method,
allows one to tackle problems that lie beyond the reach of MaxEnt \cite%
{Caticha03}.

As an illustration we briefly consider in section 4 an application in
statistical physics. Specifically, we \emph{formulate} the question: what is
the value of a certain macroscopic quantity? We obtain a full probability
distribution and not merely an estimate of the answer.

\section{What is a question?}

Our uncertainty about a variable $x$ is described by a probability
distribution $q(x)$. The distribution $q(x)$ is a faithful representation of
the information we have about $x$, and conversely it also represents the
information that we do not have and that we \emph{presumably desire to obtain%
}. Indeed, to the probability distribution that reflects lack of a certain
piece of information it is natural to associate the question that requests
it. Conversely, to the question that requests a certain information one can
associate the probability distribution that represents the rational beliefs
of someone who lacks that particular piece of information. This suggests
that questions interpreted as requests for information can be identified
with the corresponding probability distributions:

\noindent \textbf{Definition PD:} \emph{A question is a probability
distribution.}

One immediate consequence of this definition is that the age-old controversy
about whether probabilities are subjective or objective or some mixture of
both carries over to the nature of questions. Another consequence is that
the rules to manipulate questions are fixed by the requirement that they be
compatible with the rules to manipulate the corresponding probability
distributions -- the sum and the product rules. Perhaps the main advantage
is that a properly formulated PD question explicitly incorporates all we
know and makes estimating the answer straightforward (at least in
principle); short of acquiring more information this is the best we can do.

On the other hand the Cox-Knuth (CK) definition identifies questions with
the set of answers:

\noindent \textbf{Definition CK:} \emph{A question is a set of statements
that includes all statements that qualify as answers plus all other
statements that imply those answers.}

The definitions CK and PD are sufficiently different that one might wonder
whether the two resulting theories of inquiry are at all compatible.
Assertions, that is, answers, form a Boolean lattice that is partially
ordered by the relation of implication and on which one can define
probabilities as degrees of partial implication. The foundation of the CK
approach to logical inquiry is that the corresponding questions also form a
lattice, which, however, is not Boolean (the notion of the complement or
negation of a question is not, in general, defined \cite{Knuth02}).

The PD definition requires that the set of answers $\{x\}$ be specified and
the values of $x$ must be exhaustive and mutually exclusive. Such sets of
answers lead to what in the CK approach are called \textquotedblleft
partition\textquotedblright\ questions. The CK definition, however, also
allows other kinds of questions that are strange and do not quite conform
with our intuition of what questions ought to be. These strange questions
include \textquotedblleft vain\textquotedblright\ questions which having no
true answers are clearly the \textquotedblleft wrong\textquotedblright\
question to ask. For example, the question `Is the book green or blue?' is
vain when the book happens to be red; `Is the electron a particle or a
wave?' is another vain question because the true answer is `neither'.
Questions that do have a true answer are called `real' and conform to our
intuition of what a \textquotedblleft correct\textquotedblright\ question
is. Also strange are the so-called \textquotedblleft
ideal\textquotedblright\ questions which have only one answer. For example,
the set of answers that defines the question `What color is Napoleon's white
horse?' is limited to the single element `white' and it is not clear why one
would bother to ask such a \textquotedblleft dumb\textquotedblright\
question in the first place. Remarkably, despite appearances, these
\textquotedblleft wrong\textquotedblright\ and \textquotedblleft
dumb\textquotedblright\ questions actually play useful roles within the CK
approach to logical inquiry.

From the perspective of the previous paragraph PD questions are special
cases of CK questions. But this is not the end of the story. To the
exhaustive and mutually exclusive set $\{x\}$, which defines a single unique
CK real question, one can associate an infinity of probability
distributions. Therefore the single CK question can be resolved into an
infinity of different PD questions. From this second perspective it is the
PD questions that constitute the generalization.

These brief remarks suggest that the two approaches are complementary rather
than contradictory; they explore different aspects of the logic of inquiry.
I suspect that a more complete theory of inquiry will eventually be
constructed by a fusion of concepts from both.

\section{Entropy as a measure of relevance}

Given a question $Q(x)$ the ultimate goal is to find its answer and this is
not always easy. One of the strategies one can follow is to replace $Q(x)$,
which we will call the \emph{prior} question, by another presumably easier
question $q(x)$. This might not completely resolve the prior question but,
at least, it may represent some progress. The problem is to select from a
given family of trial \textquotedblleft easy\textquotedblright\ questions
that which is most \textquotedblleft relevant\textquotedblright\ to
resolving the prior question. Typically the family of trial questions will
be defined by suitable constraints on the corresponding probability
distributions.

We want to rank the questions within the trial family according to their
relevance, then it will be easy to pick the most relevant one. The ranking
must be transitive: if, relative to the prior question $Q$, question $q_{1}$
is more relevant than question $q_{2}$, and $q_{2}$ is itself more relevant
than $q_{3}$, then $q_{1}$ is more relevant than $q_{3}$. Such transitive
rankings can be implemented by assigning to each question $q$ a real number $%
S[q,Q]$ and, therefore, it is natural to \textquotedblleft
measure\textquotedblright\ relevance by real numbers.

Next, we define the general theory of relevance, that is, the functional
form of $S[q,Q]$, by invoking the seemingly trivial but nevertheless
fundamental inductive principle that `\emph{If a general theory exists, it
must apply to special cases.}' \cite{Skilling88}. The idea is simple: If the
most relevant distribution happens to be known in a special case, then this
knowledge can be used to constrain the form of $S[q,Q]$. A sufficient number
of such constraints can lead to the complete specification of $S[q,Q]$.
Unfortunately, it is quite possible that a completely general theory does
not exist, that $S[q,Q]$ might be overspecified by too many special cases
and that there is no $S[q,Q]$ that simultaneously reproduces them all.

The known special cases, called the \textquotedblleft
axioms\textquotedblright , play the crucial role of defining which general
theory is being constructed; they define what we mean by `relevance'.

To motivate the choice of axioms we consider first an extreme example (a
special case of axiom 1 below): Within the unconstrained family of all
possible questions $\{q\}$, the question $q_{Q}$ that is most relevant to
answering $Q$ is $Q$ itself, $q_{Q}=Q$. Indeed, the answer to $q_{Q}$ tells
us precisely the answer to $Q$. The axioms below are elaborations of the
basic idea that one question is very \textquotedblleft
relevant\textquotedblright\ to another when both are essentially requesting
the same information, that is, when the corresponding distributions are very
\textquotedblleft close\textquotedblright\ to each other.

Three axioms, a brief justification and their consequences for the form of $%
S[q,Q]$ are given below (detailed proofs appear in \cite{Caticha04}).

\textbf{Axiom 1: Locality}. \emph{Local constraints have local effects.} If
the constraints that define a family of trial questions do not refer to a
certain domain $D\subseteq \{x\}$, then the selected most relevant question $%
q_{Q}(x)$ and the prior question $Q(x)$ should, within $D$, request the same
information, that is, the corresponding conditional probabilities should
coincide, $q_{Q}(x|D)=Q(x|D)$. The consequence of the axiom is that
non-overlapping domains of $x$ contribute additively to the relevance: $%
S[q,Q]=\int dx\,F_{Q}(q(x))$ where $F_{Q}$ is some unknown function.

\textbf{Axiom 2: Coordinate invariance.} \emph{The ranking according to
relevance does not depend on the system of coordinates. }The coordinates $x$
that label the possible answers are arbitrary; they carry no information.
The consequence of this axiom is that $S[q,Q]=\int dx\,q(x)f(q(x)/m(x))$
involves coordinate invariants such as $dx\,q(x)$ and $q(x)/m(x)$, where $%
m(x)$ is a density, and both functions $m(x)$ and $f$ are, at this point,
still to be determined.

Next we determine $m(x)$ using the special instance of the locality axiom
that was mentioned earlier: we allow the domain $D$ to extend over the whole
space $\{x\}$ so that within the unconstrained family of all possible
questions $\{q\}$, the question that is most relevant to $Q$ is $Q$ itself.
But maximizing $S[q,Q]$ subject to no constraints gives $q_{Q}(x)\propto
m(x) $, and therefore $m(x)\propto Q(x)$. Up to normalization $m(x)$ is $Q(x)
$.

\textbf{Axiom 3:\ Subsystem independence}. \emph{When a system is composed
of subsystems that are believed to be independent it should not matter
whether our inquiries treat them separately or jointly.} Specifically, let
the possible states of a composite system be described by $(x_{1},x_{2})$.
Suppose that when we treat the subsystems separately we decide that the
question about subsystem 1 that is most relevant to the prior question $%
Q_{1}(x_{1})$ is $q_{Q_{1}}(x_{1})$ and, similarly, that for subsystem 2 the
question that is most relevant to $Q_{2}(x_{2})$ is $q_{Q_{2}}(x_{2})$. On
the other hand, when we treat the subsystems jointly the prior question is $%
Q(x_{1},x_{2})=Q_{1}(x_{1})Q_{2}(x_{2})$. This question does not request
information about correlations because we already believe the subsystems are
independent. We seek the most relevant distribution $q_{Q}(x_{1},x_{2})$
within a family $\{q(x_{1},x_{2})\}$. Axiom 3 states that the selected $%
q_{Q}(x_{1},x_{2})$ should also not request information about correlations
and that it should be $q_{Q_{1}}(x_{1})q_{Q_{2}}(x_{2})$ unless this product
distribution is explicitly excluded from the family $\{q(x_{1},x_{2})\}$.

The consequence of axiom 3 is that the function $f$ is restricted to be a
logarithm. (The fact that the logarithm applies also when the subsystems are
not independent follows from our inductive hypothesis that the ranking
scheme has general applicability.)

The overall consequence of these three axioms is that questions $q(x)$
should be ranked relative to the central question $Q(y)$ according to the
(relative) entropy of the corresponding distributions, 
\begin{equation}
S[q,Q]=-\int dx\,q(x)\log \frac{q(x)}{Q(x)}.  \label{S[q,Q]}
\end{equation}%
This derivation has singled out the entropy $S[p,Q]$ as \emph{the unique
criterion for ranking according to relevance}. Other criteria may be useful
for other purposes but they are not a generalization from the simple cases
described in the axioms.

\section{Questions and relevance in statistical physics}

Let the microstates of a physical system be denoted by $x$; let $m(x)dx$ be
the number of microstates in the range $dx$; and let the system be in
equilibrium at temperature $T$, 
\begin{equation}
P(x)=\frac{1}{Z}\,m(x)\,e^{-\beta H(x)},  \label{P(x)}
\end{equation}%
where $H(x)$ is the Hamiltonian, $\beta =1/kT$, and $Z=\int dx\,m\,e^{-\beta
H}$. On the basis of the information in $P(x)$ many quantities can be
estimated accurately but, when it happens that estimates of some other
quantities are not satisfactory we might suspect that not all relevant
information has been taken into account.

Let us assume that the information that we should have taken into account is
the value of a macroscopic variable $A$. (With suitable notation changes we
could easily consider several variables.) From a microscopic point of view $%
A $ is interpreted as the expected value of some micro-variable $a(x)$, $A%
\overset{\limfunc{def}}{=}\left\langle a\right\rangle $, and the
distribution that incorporates this information and no more is

\begin{equation}
q(x|A)=\frac{1}{Z(\lambda )}\,m(x)\,e^{-\beta H(x)-\lambda a(x)},
\label{q(x|A)}
\end{equation}%
where $Z(\lambda )$ is the appropriate partition function and the Lagrange
multiplier $\lambda $ is determined from $\partial \log Z/\partial \lambda
=-A$.

Our goal is to \emph{formulate} the question `What is $A$?'. Of course, it
is straightforward to find an answer on the basis of the information
codified into $P(x)$; just average over $P$ to get $A\approx \left\langle
a\right\rangle _{P}$. This is the best we can do with the information in $P$
but here we deal with a situation where we have the \emph{additional}
information that the \textquotedblleft true\textquotedblright\ (\emph{i.e.}
a much better) distribution is a member of the family $q(x|A)$. \mathstrut
By \emph{formulating} the question `What is $A$?' we mean to be very
explicit about what information is being requested and about the information
that is already available; we want a probability distribution $q(A)$.

In $P(x)$ we were originally asking `What is $x$?'; now we also ask `What is 
$A$?'. The universe of discourse is not the set of microstates $\{x\}$, but
the larger set $\{A,x\}$; the questions we now ask are joint distributions.
Of all questions $q(A,x)$ within a certain family we seek that which is
closest to our original prior question $Q(A,x)$ determined by two
requirements: first, that $\int Q(A,x)dA=P(x)$, and second, that when we
know absolutely nothing about $A$, $Q(A,x)$ must be a product $\mu (A)P(x)$
with $\mu (A)$ chosen as uniform as possible. The question of what do we
mean by `uniform' is settled by supplying the additional information that to
each $A$ there corresponds a probability distribution $q(x|A)$, eq.(\ref%
{q(x|A)}). Thus, there is a natural measure $\mu (A)=\gamma ^{1/2}(A)$ in
the space of $A$s which is given by the Fisher-Rao metric between the
corresponding probability distributions $q(x|A)$, 
\begin{equation}
\gamma (A)=\int dx\,q(x|A)\left[ \frac{\partial \log q(x|A)}{\partial A}%
\right] ^{2}.
\end{equation}

We are now ready to take the final step: to find the question $q(A,x)$ that
is closest, most relevant to the prior question $Q(A,x)=\gamma ^{1/2}(A)P(x)$
we maximize 
\begin{equation}
S[q,Q]=-\int dA~dx~q(A,x)\log \frac{q(A,x)}{\gamma ^{1/2}(A)P(x)}
\end{equation}%
subject to the constraint that $q(A,x)$ be normalized and that it be of the
form $q(A,x)=q(A)q(x|A)$ with $q(x|A)$ given by eq.(\ref{q(x|A)}). Varying
over $q(A)$ gives 
\begin{equation}
q(A)dA\propto e^{S(A)}\gamma ^{1/2}(A)dA\quad \text{where\quad }S(A)=-\int
dx~q(x|A)\log \frac{q(x|A)}{P(x)}~.  \label{q(A)}
\end{equation}%
\mathstrut Note that the density $\exp S(A)$ is a scalar function and the
presence of the Jacobian factor $\gamma ^{1/2}(A)$ makes Eq.(\ref{q(A)})
manifestly invariant under changes of the coordinates $A$.

The distribution $q(A)$ is the question we sought to formulate. Once the
question is formulated we can estimate an answer. \mathstrut The most
probable $A$ (the value that maximizes the probability per unit volume) is
such that $q(x|A)$ is closest to requesting the same information as the
original $P(x)$; it maximizes the relevance $S(A)$ of the question $q(x|A)$
relative to the prior question $P(x)$. Eq.(\ref{q(A)}) also tells us the
degree to which values of $A$ away from the maximum are ruled out by the
available information.

\section{Conclusion}

In this paper the logic of inquiry has been explored from a point of view
that captures an aspect of what we mean by a question -- a question is a
request for information -- that had not been previously addressed within the
Cox-Knuth approach.

First we asked `What is \emph{a} question?' and then we took a first step in
addressing the issue of `What is \emph{the} question?' at least in the
restricted context of selecting the question within a family that is most
relevant to answering another \textquotedblleft prior\textquotedblright\
question.

It is to be expected that the ideas explored here should apply whenever we
are confronted with deciding what is the most relevant question to ask.
Obvious examples include the optimal design of experiments and when
selecting which variables are most relevant to the description of a certain
phenomenon. At the very least it is clear that much remains to be done, both
in terms of seeking a closer integration with the Cox-Knuth approach, in
pursuing applications, and ultimately in seeking a deeper understanding of
both `What is a question?' and `What is the question?'.

\noindent \textbf{Acknowledgments- }I am indebted to K. Knuth, R. Fry, and
C.-Y. Tseng for many insightful remarks and valuable discussions.


\begin{thebibliography}{9}
\bibitem{Cox61} R. T. Cox, \emph{The Algebra of Probable Inference} (Johns
Hopkins, Baltimore, 1961).

\bibitem{Cox79} R. T. Cox, \textquotedblleft Of Inference and Inquiry, an
Essay in Inductive Logic\textquotedblright , p. 119-167 in \emph{The Maximum
Entropy Formalism}, R. D. Levine and M. Tribus (eds.) (MIT Press, Cambridge,
1979).

\bibitem{Fry95} R. L. Fry, \textquotedblleft Observer-participant models of
neural processing\textquotedblright\ IEEE Trans. Neural Networks, \textbf{6}%
, 918 (1995); R. L. Fry, \textquotedblleft The Engineering of Cybernetic
Systems\textquotedblright\ in \emph{Bayesian Inference and Maximum Entropy
Methods in Science and Engineering}, R. L. Fry (ed.), AIP Conf. Proc. 
\textbf{617}, 497 (2002).

\bibitem{Knuth02} K. H. Knuth, \textquotedblleft What is a
question?\textquotedblright\ in \emph{Bayesian Inference and Maximum Entropy
Methods in Science and Engineering}, C. Williams (ed.), AIP Conf. Proc. 
\textbf{659}, 227 (2002); K. H. Knuth, \textquotedblleft Deriving laws from
ordering relations\textquotedblright\ in \emph{Bayesian Inference and
Maximum Entropy Methods in Science and Engineering}, G. Erickson and Y. Zhai
(eds.), AIP Conf. Proc. \textbf{707}, 204 (2004), K. H. Knuth,
\textquotedblleft Lattice duality: the origin of probability and
entropy\textquotedblright\ to appear in Neurocomputing (2004).

\bibitem{Caticha04} A. Caticha, \textquotedblleft Relative Entropy and
Inductive Inference,\textquotedblright\ in \emph{Bayesian Inference and
Maximum Entropy Methods in Science and Engineering}, G. Erickson and Y. Zhai
(eds.), AIP Conf. Proc. \textbf{707}, 75 (2004)
(arXiv.org/abs/physics/0311093).

\bibitem{ShoreJohnson80} J. E. Shore and R. W. Johnson, \textquotedblleft
Axiomatic derivation of the Principle of Maximum Entropy and the Principle
of Minimum Cross-Entropy,\textquotedblright\ IEEE Trans. Inf. Theory \textbf{%
IT-26}, 26 (1980); \textquotedblleft Properties of Cross-Entropy
Minimization,\textquotedblright\ IEEE Trans. Inf. Theory \textbf{IT-27}, 26
(1981).

\bibitem{Skilling88} J. Skilling, \textquotedblleft The Axioms of Maximum
Entropy\textquotedblright\ in \emph{Maximum-Entropy and Bayesian Methods in
Science and Engineering}, G. J. Erickson and C. R. Smith (eds.) (Kluwer,
Dordrecht, 1988); J. Skilling, \textquotedblleft Classic Maximum
Entropy\textquotedblright\ in \emph{Maximum Entropy and Bayesian Methods},
J. Skilling (ed.) (Kluwer, Dordrecht, 1989).

\bibitem{Caticha03} A. Caticha and R. Preuss, \textquotedblleft Entropic
Priors\textquotedblright\ in \emph{Bayesian Inference and Maximum Entropy
Methods in Science and Engineering}, G. Erickson and Y. Zhai (eds.), AIP
Conf. Proc. \textbf{707}, 371 (2004); and \textquotedblleft Maximum Entropy
and Bayesian Data Analysis: Entropic Prior Distributions\textquotedblright\
to appear in Phys. Rev. \textbf{E} (arXiv.org/abs/physics/0307055). C.-Y.
Tseng and A. Caticha, \textquotedblleft Maximum Entropy Approach to the
Theory of Simple Fluids\textquotedblright\ in \emph{Bayesian Inference and
Maximum Entropy Methods in Science and Engineering}, G. Erickson and Y. Zhai
(eds.), AIP Conf. Proc. \textbf{707}, 17 (2004)
(arXiv.org/abs/physics/0310746).
\end{thebibliography}
\end{document}